\documentstyle[twocolumn,epsfig]{article}
\def\pg{{\cal P}}
\def\g{\gamma}
\def\e{\varepsilon}

\begin{document}

\title{\bf Probability distribution of the conductance
at the mobility edge}

\author{%
Peter Marko\v s\\ Institute of Physics, Slovak Acad. Sci, D\'ubravsk\'a cesta 9, 842 28 Bratislava, Slovakia\\
}

\maketitle

\abstract{Distribution of the conductance $P(g)$
at the critical point of the metal-insulator transition is 
presented for three and four dimensional orthogonal systems.
The form of the distribution is discussed. 
Dimension dependence of $P(g)$ is proven. The limiting cases
$g\to\infty$ and $g\to 0$ are discussed in detail and relation
$P(g)\to 0$ in the limit  $g\to 0$ is proven
}

\medskip


\bigskip

As the conductance $g$ in disordered systems is not 
the self-averaged quantity, the knowledge of its probability distribution
is extremely important for our understanding of transport. This problem
is of special importance at the critical
point of the metal-insulator transition 
\cite{shap}.
While the distribution
of the conductance in the metallic phase 
is known to be Gaussian in  agreement with the random-matrix theory
\cite{pichard} 
and the localized regime is characterized by the log-normal distribution
of $g$
\cite{pichard},
the form of the critical distribution remains still unknown.
Among the problems which are not  solved yet we mention   e.g.
the number of parameters which characterize
distribution, the existence of huge fluctuations of the conductance,
and the form of $P(g)$ for small values of $g$.

Several attempts has been made to characterize conductance distribution
at the critical point. 
Using the Migdal - Kadanoff renormalization treatment,
huge conductance fluctuations has been predicted  in \cite{CRS}.
The same conclusion was found also in systems of 
dimension $d=2+\varepsilon$. In the limit $\varepsilon<<1$
the form of the distribution $P(g)$
was found analytically
\cite{shapiro}. However, numerical studies of  disordered 3D system
\cite{pm} indicated that it is not possible 
to generalize these analytical conclusions for realistic 3D 
systems ($\varepsilon=1$). 

The form of $P(g)$ for 2D symplectic
models was found in \cite{jp,el}.
Recently, $P(g)$ has been studied also for system in magnetic field,
both in 3D \cite{slevin} and in 2D \cite{soukoulis}. 
The main conclusion of these studies is that the symmetry of the system
influences the form of the distribution at the 
critical point more strongly than in the metallic or localized 
regime. Nevertheless, $P(g)$ is invariant with
respect to the choice of the microscopic model within the same universality
class \cite{el}.

Studies of the statistics of the conductance have their counterpart
in the analysis of the level statistics
$s=E_{i+1}-E_i$ of the eigenvalues of Hamiltonian
\cite{shkl}. The critical distribution $P(s)$
is also the 
subject of intensive studies within last years \cite{zk}. 
In particular, its dependence on the symmetry
\cite{isa},
and dimension
\cite{zkk}
have been studied  numerically.

In this Letter we present new numerical data 
for the 3D and 4D  Anderson model (orthogonal
ensembles).  Although data prove the dimension dependence of the distribution, their  enables us to 
discuss the common features of the critical distribution. In particular, we  
prove  that $P(g)$ decreases more quickly than exponentially
for large $g$. This assures that there are no huge  
fluctuations of the  conductance, discussed in \cite{CRS}. We prove also
that $P(g)\to 0$ in the limit of $g\to 0$.

We calculated the conductance as
\begin{equation}\label{g}
g={\rm Tr}~t^{\dag} t=\sum \cosh^{-2}(z_i/2)
\end{equation}
where quantities $z_i$
determine eigenvalues of the transmission matrix $t^\dag t$. 
Details of the method have been published elsewhere \cite{pm}. 
For a given system size $L$, the probability distribution of $g$
has been calculated from an ensemble of $N_{\rm stat}$ samples.
The list of used ensembles together with mean and variances of $g$
are given in Table 1. 

The last column of Table 1 presents parameter $\langle z_1\rangle$,
which   corresponds to the
parameter $\Lambda$ introduced in the finite size scaling theory
by MacKinnon and Kramer
\cite{mackinnon} as $\langle z_1\rangle={2L_t\over L\Lambda}$ in the
quasi-one dimensional limit $L^{d-1}\times L_t$, $L_t>>L$.
When neglecting the smallest system size, our data confirm the
$L$-invariance of $\langle z_1\rangle$ as well as of 
$\langle g\rangle$ and $\langle\log g \rangle$ and their standard deviations.
Owing to higher critical disorder, $\langle z_1\rangle$ is larger in 4D 
than in 3D. 
This guarantees  that the finite-size effects disappear more quickly in 4D.
Therefore, in spite of the fact that computer facilities limited the 
system size to $L\le 8$ for $d=4$,
obtained data provide us with the relevant information
about all parameters of interest. 

We presented in Table 1 both mean values of $g$ and $\log g$ to underline
the common features of 3D and 4D distribution: the variance of $\log g$ is
of order of its mean value. This relation is typical for localized state.
On the other hand, standard deviation of $g$ is also $\sim \langle g\rangle$.
Its value for 3D samples, 0.334, is smaller than the same quantity
calculated for 3D in the metallic regime \cite{pm}.

\begin{figure}[t]
\begin{center}
\epsfig{file=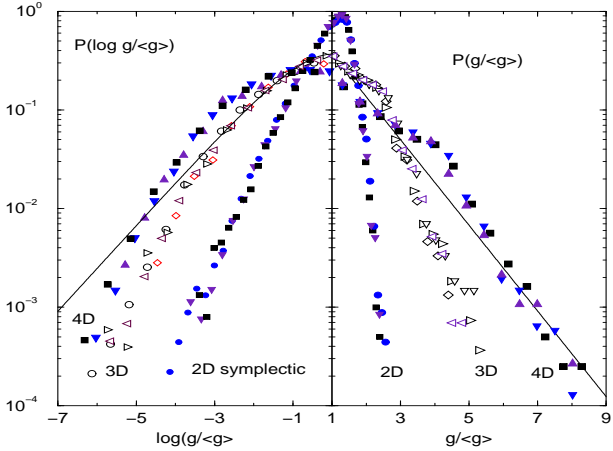,width=8cm,height=6cm}
\end{center}
\caption{Probability distribution of $\log g/\langle g\rangle$  (left)
and $g/\langle g\rangle$ (right) for 4D (full symbols) and
3D (open symbols) Anderson model. For comparison,
we plot  also data for 2D (symplectic) Ando model. The last exhibit the best
convergence for both small and large values of $g$.
For meaning of symbols,
see Table 1.  Solid line is Poisson distribution $P(g)=\exp -g/\langle g\rangle$. }
\end{figure}

\begin{table}[t]
{\small
\caption{Review of ensembles studied in the present work. $L$-size
of the $d$-dimensional cube, $N_{\rm stat}$: number of samples in a given ensemble, var $g=\langle g\rangle^2-\langle g^2\rangle$, $\langle z_1\rangle$ is mean of the smallest of $z$'s. 
Data for 3D AM are in good agreement  with \cite{soukoulis} (up to the
spin degeneracy factor 2).  }
\begin{center}
\begin{tabular}{rcrccccc}
\hline
$L$ &  & $N_{\rm stat}$ & $\langle g\rangle$  &  $\sqrt{{\rm var} g}$ & $\langle\log g\rangle$ & var $\log g$ & $\langle z_1\rangle$\\
\hline
\multicolumn{7}{c}{3D Anderson model: $W_c\approx 16.5$}\\
6  &  $\circ$            & 20.000   &   0.375   &      0.324   & -1.481& 1.344 &     2.901\\
8  &  $\lhd$             &20.000   &   0.400   &      0.333   &  -1.384& 1.251 &    2.803 \\
10 &  $\rhd$             & 10.000   &   0.410   &      0.337   &  -1.347&  1.229 &    2.770\\
12 &  $\diamondsuit$     & 5.000   &   0.421   &      0.340   &  -1.302& 1.199&    2.724\\
14 &  $\bigtriangleup$   & 2.500   &   0.416   &      0.338   &  -1.306  & 1.122& 2.725\\
18  & $\bigtriangledown$ &   500   &   0.418   &      0.329   &  -1.279  & 1.083&   2.717\\
\hline
\multicolumn{7}{c}{4D Anderson model: $W_c\approx 34.5$  \cite{zk}}\\
4   &         &   22.000      &  0.190  &    0.247 & -2.569 & 2.301 &  4.130\\
5   &  $\bigtriangledown$  &    30.000      &  0.229  &    0.270 &-2.275 & 2.006 &   3.838\\
6   &  $\Box$           &    15.000      &  0.225  &    0.269 & -2.291 & 2.054    & 3.852\\
7   &  $\bigtriangleup$      & 7.000       & 0.239      &  0.275        & -2.193 & 1.971 &  3.748  \\
8   &     &    200  &  0.227      &    0.274   & -2.188    &  1.692  &  3.790  \\
\hline
%\multicolumn{7}{c}{2D Ando model: $W_c\approx 5.75$}\\
%10  &  $\circ$           & 20.000  &  1.481   &  0.640  & 0.261  & 0.337   & 1.52   \\
%20  &  $\bigtriangledown$& 10.000  &  1.488   &  0.619  & 0.275 & 0.312  & 1.47  \\
%30  &  $\Box$            & 20.000  &  1.485   &  0.612  & 0.275  & 0.307   & 1.46   \\
%60  &                    &  4.000  &  1.509   &  0.602  & 0.299  & 0.287   & 1.416\\
\end{tabular}
\end{center}
}
\end{table}
Numerical data for $P(g)$  are presented in Figure 1. 
They  confirm that 
the  critical distribution of $g$ is system-size independent, in agreement
with previous studies. 
Fig. 1. shows   also that $P(g)$ depends on the
dimension of the system within the same symmetry class.
Although the distribution has the same form for 3D and
4D ensembles, 
it becomes  broader for higher  $d$: the probability to find
$g<<\langle g\rangle$ or $g>>\langle g \rangle$ growths with 
dimension. This is due to higher critical 
disorder, which causes that electronic state  
posses more features of the localized state than that of the
metallic one (remaining  critical).  
This is  in agreement with studies of the
level statistics in 4D \cite{zk}.

The  small- $g$ behavior of $P(g)$
can be estimated  from Fig 1.
Instead of $P(g/\langle g\rangle)$, we plot in the left side of Fig. 1. 
the distribution $\pg(\g)$ of $\g=\log g/\langle
g\rangle$. Evidently, $\log\pg(\g)=\g+\log P(\exp\g)$.
Therefore, an assumption  $P(g=0)=c\ne 0$, implies $\pg=\g+\log c$ for
$\g\to -\infty$. 

Fig 1. shows clearly that $\log\pg(\g)$ decreases
more quickly than $\g$ for {\sl all} ensembles we consider.
This guarantees that $P(g)\to 0$ as $g\to 0$. Let us note that
it is almost impossible to obtain last result form the studies of
$P(g)$ on the linear scale
\cite{soukou}.

\begin{figure}[t]
\begin{center}
\epsfig{file=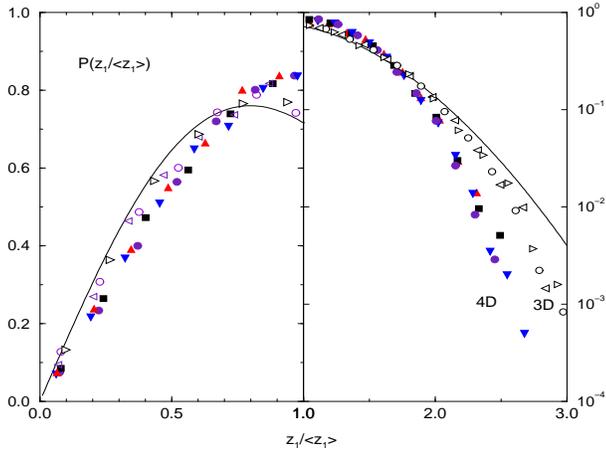,width=8cm,height=6cm}
\end{center}
\caption{Probability distribution of (normalized) $z_1$ for 4D and
3D Anderson model. Solid line is Wigner surmises 
$P_W(z)={\pi\over 2}z\exp[-{\pi\over 4}z^2]$. For the mean
 value $\langle z_1\rangle$ see Table 1.}
\end{figure}

The small- $g$ behavior of $P(g)$ is easy to estimate also from
the distribution $P(z_1)$ of the smallest parameter $z_1$. 
Indeed, small values of  $g$ require large values of  $z_1.$ Neglecting 
contributions of other  channels, we have
\begin{equation}\label{gsmall}
{1\over 2\varepsilon}\int_0^{2\varepsilon} 
P(g)dg={1\over 2\varepsilon}\int_{\tilde{z_1}}^\infty P(z_1)dz_1
\end{equation}
with
$\varepsilon=\exp -\tilde{z_1}$. In the limit $\varepsilon\to 0$ the integral
on the LHS reads $\sim P(g)$, $g=\varepsilon$. RHS could be found analytically
for special form of $P(z_1)$. In particular, for Wigner surmises 
$P(z_1)=\pi/2\langle z_1\rangle^2\times z_1\exp(-\pi/4\times 
[z_1/\langle z_1\rangle]^2)$ 
we obtain that  $P(g)\sim  g^{-1-{\rm const}\times\log g}/2$ 
with const = ${\pi\over 4\langle z_1\rangle^2}$. Consequently,
$P(g=0)=0$. 
Fig. 2. assures that $P(z_1)$ decreases 
more quickly than Wigner surmise for large $z_1$ in orthogonal ensemble 
for both 3D and 4D systems.
This assures that $P(g)\to 0$ as $g\to 0$.

Linear behavior  of the distribution $P(z_1)$ for small $z_1$  (see left side of Fig. 2)
guarantees nonzero probability that the first channel is fully open.  Indeed,
if $P(z_1)\sim C\times z_1$ for $z_1\to 0$, then 
the probability, that the first channel contribution to the conductance, 
$g_1=1/\cosh^2(z_1)$, equals to 1, is  $C$.
This explains the origin 
of the characteristic bump in the distribution $P(g)$ for $g=1$.
In Fig. 1, the bump is clearly visible for both 3D and 4D systems.

%Data in Fig. 1.   present strong argument {\sl against} the existence
%of the large fluctuations of $g$ at the critical point as were discussed in
%\cite{CRS,shapiro}.
%In fact, we never obtained $g>2.13$ (in 3D) or $g>2.06$ (in 4D).   
Fig. 1 (right) confirms that $P(g)$ decreases more quickly than exponentially
for large $g$.   This is easy
to understand on the basis of the analysis of the statistics of
$z$'s presented in \cite{pm}.
\begin{figure}[t]
\begin{center}
\epsfig{file=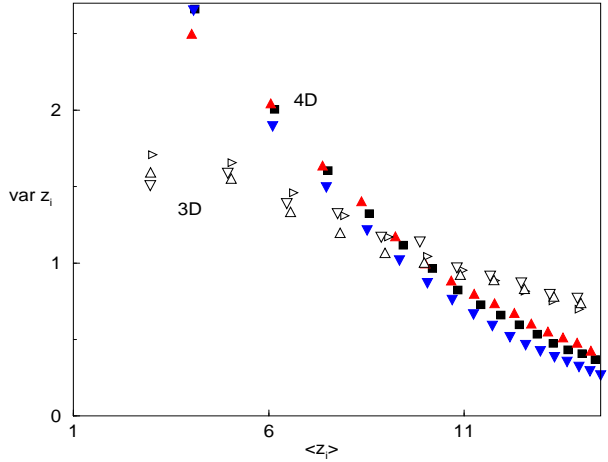,width=8cm,height=6cm}
\end{center}
\caption{Var $z_i$ as a function of $\langle z_i\rangle $ (only for $\langle z\rangle<15)$ for 3D (open symbols) and 4D (full symbols) orthogonal systems. Note
the system-size invariance of presented parameters (at least for $i\le L$). } 
\end{figure}
Fig 3. shows mean values and variances of some smallest $z$'s for both
3D and 4D system. Evidently, $\langle z_i\rangle\sim{\cal O}(1)$ and variances
var $z_i$ decreases quickly with index $i$. Consequently, the contribution to the conductance
from the second  (higher) channel is, due to (\ref{g}), small (negligible).
To estimate this contribution, we note that
all higher $z_i$, $i\ge 2$, are normally distributed.
\cite{pm}.
Their mean and variances has been estimated as
$\langle z_i\rangle\sim\langle z_1\rangle\times i^{1/(d-1)}$
and var $z_i\sim\langle z_i\rangle^{-(d-2)}$
\cite{jpcm}. Although this result holds only in the quasi-one dimensional
limit, where the mutual correlations of $z$'s are negligible, they serve
as a good quantitative estimation also for true $d$-dimensional cubes.
As i is seen in Figure 3, this agreement is better for 4D than for  3D,.
Then,  the probability to find $g\approx n$ is less than
$\exp[-\langle z_n\rangle/2{\rm var}z_n]\sim\exp[-{\rm const}\times n^{d/(d-1)}]$
and
\begin{equation}\label{glarge}
P(g)\sim \exp -{\rm const}\times g^{d/(d-1)}.
\end{equation}

We conclude that presented numerical data  
for 3D and 4D  Anderson model  prove the 
system size invariance of the conductance distribution at the critical
point. 
Although the distribution depends on the dimension and symmetry 
of the system, we found its common features, namely 
exponential decrease of $P(g)$ for $g>1$, and
a decrease of $P(g)$ to zero for $g=0$. We show that the form of $P(g)$
can be analyzed on the basis of the statistics of parameters $z$
introduced by relation (1). This analysis is more simple for higher
dimension, where the statistical correlations of $z$s are supposed to be
less important.

\medskip

\noindent Acknowledgement This work has been supported by Slovak Grant Agency,
Grant n. 2/4109/98.


\medskip





\begin{thebibliography}{9}
\bibitem{shap} B. Shapiro, Phys. Rev. Lett. {\bf 65} 1510 (1990)
\bibitem{pichard} J.-L. Pichard, Quantum Coherence in Mesoscopic Systems,
edited by B. Kramer, NATO ASI Ser. B {\bf 254} (New York: Plenum) p. 369 (1991)
\bibitem{CRS} A. Cohen, Y.  Roth and B. Shapiro, Phys. Rev. B {\bf 38} 12125 (1988)
\bibitem{shapiro} A. Cohen and B. Shapiro, Int. J. Mod. Phys. B {\bf 6}, 1243 (1992)
\bibitem{pm} P. Marko\v s and B. Kramer, Phil. Mag. B. {\bf 68}, 357 (1993).  
\bibitem{jp} P. Marko\v s, J. Phys. I, {\bf 4}, 551 (1994).
\bibitem{el} P. Marko\v s, Europhys. Lett. {\bf 26}, 431 (1994)
\bibitem{slevin} K. Slevin and T. Ohtsuki, Phys. Rev. Lett. {\bf 78}, 4083 (1987).
\bibitem{soukoulis} Xiashoa Wang, Quiming Li, C.M. Soukoulis, cond-mat/9803356
\bibitem{soukou} C.M. Soukoulis, Xiaosha Wang, Qiming Li and M.M. Sigalas, 
comment to PRL (cond-mat/987410).
\bibitem{shkl} B. I. Shklovskii, B. Shapiro, B.R. Sears, P. Lambrianides, H.B. Shore, Phys. Rev. B {\bf 47}, 11487 (1993)
\bibitem{zk} I.Kh. Zharekeshev and B. Kramer, Phys. Rev. Lett. {\bf 79}, 717 (1997)
\bibitem{isa} M. Batsch, L. Schweitzer, I. Kh. Zharekeshev, B. Kramer, Phys. Rev. Lett. {\bf 77}, 1552 (1996)
%\bibitem{milde} F. Milde and R.A. R\"omer, to appear in Annalen der Physik (1999)
\bibitem{zkk} I.Kh. Zharekeshev and B. Kramer, to appear in Ann.Phys. (cond-matt/9810286)
\bibitem{mackinnon} A. MacKinnon and B. Kramer, Phys. Rev. Lett. {\bf 47} 1546 (1981)
\bibitem{jpcm} P. Marko\v s, J. Phys.: Condens. Matt. {\bf 7}, 8361 (1995)
\bibitem{jpa}  P. Marko\v s, J. Phys.A:  Math. Gen.     {\bf 30} 3441 (1997)



\end{thebibliography}
\end{document}